\documentclass[cite,figures]{epl}
\usepackage{amsmath}
\usepackage{amssymb}
\usepackage{epsfig,color}

\shorttitle{Conventional Superconductivity in Fe-based Pnictides}
\shortauthor{{\sc M.L.\ Kuli\'{c}} \etal}
\institute{
\inst{1}Goethe-Universit\"{a}t Frankfurt, Theor.\ Physik,
D-60054 Frankfurt/Main, Germany \\
\inst{2} Leibniz-Inst.\ f.\
Festk\"orper- u.\ Werkstoffforschung Dresden (IFW Dresden), Germany\\
\inst{3}Max-Planck-Institut f\"{u}r Festk\"{o}rperphysik, D-70569
Stuttgart, Germany\\}
\input{tcilatex}
\begin{document}

\title{Conventional Superconductivity in Fe-Based Pnictides: the Relevance
of Intra-Band Electron-Boson Scattering}
\author{\textsc{M.L.\ Kuli\'{c}}\inst{1} \and \textsc{S.-L.\ Drechsler}%
\inst{2} \and \textsc{O.V.\ Dolgov}\inst{3}}
\maketitle

\begin{abstract}
{\footnotesize {\textcolor{black}{Various r}ecent experimental %
\textcolor{black}{data} 
and \textcolor{black}{especially the} large Fe-isotope effect
point against unconventional pairing\textcolor{black}{s}, since
the large intra-band impurity scattering is strongly pair-breaking
for \textcolor{black}{them.}
The strength of the inter-band impurity scattering in some single crystals
may be strong and probably beyond the Born scattering limit. In that case
the proposed $s_{\pm }$ pairing (\textcolor{black}{hole({\it h})}- and electron%
\textcolor{black}{({\it el}}-gap are of opposite signs) is
suppressed but
\textcolor{black}{possibly} not \textcolor{black}{completely}
destroyed. The
\textcolor{black}{data} imply that the intra-band pairing in the %
\textcolor{black}{\it h-} 
and \textcolor{black}{in the {\it el}-}band, 
which are inevitably due to some nonmagnetic \textcolor{black}{\it
el-
boson} interaction (EBI), must be taken into account. EBI is 
\textcolor{black}{e}ither \textcolor{black}{due t}o phonons (EPI)
or
\textcolor{black}{possibly due} to excitons (EEI), or both are
simultaneously
operative. We discuss their interplay briefly. 
\textcolor{black}{The large Fe-isotope effect favors the}
EPI and the $s_{+}$ pairing (the \textcolor{black}{{\it h}}- and %
\textcolor{black}{{\it el}-}gap\textcolor{black}{s} 
are in-phase).}}
\end{abstract}

\section{\textbf{Introduction}}


Recently, superconductivity (SC) with a relatively high critical temperature
$T_{c}$ was discovered in several families of Fe-pnictide 
\textcolor{black}{m}aterials. In the electron
\textcolor{black}{({\it el})}
doped ($1111$) system \textcolor{black}{LaFeAsO$_{1-x}$F$_{x}$}  one has $%
T_{c}\approx 26$ $K$ (and $43$ $K$ at high pressure) \cite{LaFeAsOF}. By
replacing La by \textcolor{black}{rare earths} 
$T_{c}$ is higher with the present record of $T_{c}\approx 55$ $K$ in %
\textcolor{black}{SmFeAsO$_{1-x}$F$_{x}$} \cite{SmFeAsO} and
$T_{c}\approx 56$
K in Sr$_{1-x}$Sm$_{x}$FeAsF \cite{wu}, ect. In the hole \textcolor{black}{({%
\it h})} doped ($122$) system %
\textcolor{black}{Ba$_{0.6}$K$_{0.4}$Fe$_{2}$As$_{2}$}  one has
$T_{c}=38$ $K$ \cite{Johrendt}. In most of the parent compounds
the SDW-type magnetic ordering occurs at $T_{SDW}\sim 150$ $K$,
which is suppressed either by
\textcolor{black}{{\it el-} or {\it h}-}doping at high pressur%
\textcolor{black}{e.} 
The vicinity of these systems to an antiferromagnetic
\textcolor{black}{(AFM)}
phase inspired pairing models based on spin fluctuations which are \textit{%
repulsive }in the $s$-wave channel. This situation resembles the one in
high-temperature superconductors (HTSC), where due to the vicinity %
\textcolor{black}{of} 
an \textcolor{black}{AFM phase} pairing 
\textcolor{black}{scenarios} based on the Hubbard and $t$-$J$
models have been
proposed. However, it turns out that by interpreting experimental %
\textcolor{black}{data of} 
HTSC in the framework of the Eliashberg\textcolor{black}{-}theory%
\textcolor{black}{,} phonons \textcolor{black}{might be} important
for the pairing mechanism of cuprates, giving rise to a large
\textcolor{black}{\it
el-phonon} coupling constant $1\leq \lambda _{ep}\lesssim 3.5$, while %
\textcolor{black}{the coupling due to} the spin fluctuations %
\textcolor{black}{is weak:} 
$\lambda _{sf}<0.3$ \cite{Kulic-review}.

The density functional band structure methods (DFT-LDA) applied to the
Fe-based superconductors predict at least four bands at the Fermi energy %
\textcolor{black}{$E_F$}- two 
\textcolor{black}{{\it h}-}bands around the $\Gamma $-point and
two
\textcolor{black}{{\it el-}}bands around the $M$-point
\cite{Lebegue-Singh}, which have been already observed in ARPES
\cite{epl-ARPES}.
\textcolor{black}{T}he incipient magnetism in the parent compounds
of the Fe-pnictides provoked proposals for unconventional
\textcolor{black}{SC}
either gapless or nodeless singlet or triplet pairing
\cite{Patrick-Lee}, \cite{Gorkov}. A possible excitonic-like
\textcolor{black}{mediated SC}
has been mentioned in \cite{Cvetkovic}, while \textcolor{black}{an
{\it
el-exciton} mediated} 
pairin\textcolor{black}{g} 
(EEI) \textcolor{black}{has been} 
proposed in \cite{Sawatzky}. Rather large electronic polarizability $\alpha
_{e}=(9-12)$ \AA $^{3}$ of the $As$ ions is estimated in \cite{Sawatzky},
which screens the Coulomb 
\textcolor{black}{repulsion} between 
\textcolor{black}{the charge} carriers. Some NMR measurements on
the Knight shift and the $T_{1}^{-1}$ relaxation rate were
interpreted in terms of an unconventional
(\textcolor{black}{$d$}-wave like) gapless pairing \cite{NMR}.
Based on a\textcolor{black}{n} LDA analysis of the dynamical spin
susceptibility, in \cite{Mazin-interband} the so called $s_{\pm }$
pairing
in the two-band mode\textcolor{black}{l has been proposed}.  In fact%
\textcolor{black}{,} the $s_{\pm }$ pairing scenario, where the
pairing is due to a repulsive inter-band interaction either by the
direct Coulomb one or spin fluctuations and the superconducting
gaps on the \textcolor{black}{{\it
h-}} 
and \textcolor{black}{{\it e}-} 
Fermi surfaces exhibit opposite signs, $sign(\Delta _{h})=-sign(\Delta _{e})$%
, is not quite new \cite{Aronov}. The proposed nonphononic pairing finds
support in the DFT-LDA calculations which give small EPI coupling constant $%
\lambda _{ep}$ for LaO$_{1-x}$F$_{x}$FeAs \cite{Boeri-Dolgov}, i.e.\ $%
\lambda _{ep}\approx 0.2$ and $T_{c}\sim 1$ $K$. The total coupling constant
$\lambda _{tot}=\lambda _{ep}+\lambda _{ee}...$ as derived from the zero
temperature penetration depth $\lambda _{L}(0)$ (adopted from $\mu SR$ %
\textcolor{black}{(muon spin rotation)} data \cite{luetkens}) is rather small $%
\lambda _{tot}\approx 0.5-\textcolor{black}{1}$. The plasma
frequency analysis (from reflectivity data) was performed within
the framework of a clean limit effective single-band approximation
adopting that at $T=0$ all charge carriers are in the superfluid
condensate. \cite{Drechsler-optics}.

Usually, the Fe-based superconductors are studied within the minimal
BCS-like two-band model which is proposed much earlier \cite{Suhl}, \cite%
{Moskalenko}. In the following, this model is used in conjunction with
experimental results to argue that in order to explain pairing in the
Fe-pnictides it is necessary to\textit{\ take into account the
electron-boson interaction (EBI), either EPI or EEI, or both},\textit{\ }%
which point to the presence of a conventional intra-band pairing.

\section{\textbf{Conventional EBI pairing}}

At present, there is \textit{no evidence}, neither experimental
nor theoretical, for a magnetic pairing mechanism. If the latter
is operative in the intra-band pairing it would give rise for an
unconventional order parameter. Contrary, there is numerous
experimental evidence against an unconventional pairing in
Fe-based pnictides: (\textit{i}) the presence of a large amount of
impurities and defects in \textcolor{black}{single crystalls} and
polycrystalline samples \cite{Luo-resistivity-AKFe2As2},
(\textit{ii})
rather isotropic bands and superconducting gaps in the ARPES spectra \cite%
{epl-ARPES}, (\textit{iii}) the large Fe-isotope effect with $\alpha
_{Fe}\approx 0.4$ for the substitution $^{56}Fe\rightarrow ^{54}Fe$ \cite%
{Liu-isotope}, (iv) the \textcolor{black}{\it s}-wave like
\textcolor{black}{$T$-dependence of
the} penetration depth, (v) the STM conductness with a nodeless gap \cite%
{Tesanovic-STM}, (vi) some NMR $T_{1}^{-1}$ spectra \cite{Nakai-NMR-ASF},
etc. The DFT-LDA calculations show appreciable change of the electronic
density of states due to the Fe-As based phonons \cite{Johrendt2}, which
points to the importance of EPI, while the large As polarizability may favor
excitonic pairing (EEI) \cite{Sawatzky}. These two EBI pairing mechanisms
favor conventional superconductivity in Fe-pnictides and will be discussed
briefly below.

\section{\textbf{Effect of impurities on $T_{c}$}}

The \textit{large residual resistivity} observed in some single crystals of $%
Ba_{1-x}K_{x}Fe_{2}As_{2}$ \cite{Luo-resistivity-AKFe2As2} \textit{excludes
gapless unconventional superconductivity}. Namely, the residual conductivity
$\sigma _{imp}=\rho _{imp}^{-1}$ in the minimal two-band model is given by
\cite{Allen}
\begin{equation}
\rho _{imp}^{-1}=\frac{1}{4\pi }(\frac{\omega _{pl,h}^{2}}{\Gamma _{h}^{tr}}+%
\frac{\omega _{pl,e}^{2}}{\Gamma _{e}^{tr}}),  \label{rho}
\end{equation}%
where the electron and hole \textit{transport scattering rates} are $\Gamma
_{h}^{tr}=\Gamma _{hh}^{tr}+\Gamma _{he}^{tr}$ and $\Gamma _{e}^{tr}=\Gamma
_{ee}^{tr}+\Gamma _{eh}^{tr}$. Here, the \textit{intra-band scattering }is
labelled by $hh$ and $ee$, while $he$ and $eh$ label the \textit{inter-band
scattering}. Since the density of states of hole- and electron-bands are
similar $N_{e}(0)\approx 1.4N_{h}(0)$ it is reasonable to assume that $%
\Gamma _{h}^{tr}\sim \Gamma _{e}^{tr}(\equiv \Gamma ^{tr})$ holds, where $%
\Gamma ^{tr}$ is related to the in-plane residual resistivity $\rho _{imp}$
\begin{equation}
\Gamma ^{tr}(K)\approx \frac{\pi }{2}\tilde{\omega}_{pl}^{2}(eV)\rho
_{imp}(\mu \Omega cm).  \label{Gamma}
\end{equation}%
The effective plasma frequency $\tilde{\omega}_{pl}^{2}\approx \omega
_{pl,h}^{2}+\omega _{pl,e}^{2}$ measures the total number of the conduction
carriers. The LDA calculations give an averaged plasma frequency ($\Omega
_{pl}$) in the Fe conduction plane $2$ $eV<\Omega _{pl}^{xx}\leq 3$ $eV$.
The estimated $\rho _{imp}$ in single crystals of %
\textcolor{black}{Ba$_{1-x}$K$_{x}$Fe$_{2}$As$_{2}$}
($T_{c}\approx 38$ $K$)
is in the range $\rho _{imp}=(30-50)$ $\mu \Omega cm$ \cite%
{Luo-resistivity-AKFe2As2}. Since $\tilde{\omega}_{pl}^{2}\approx (\Omega
_{pl}^{xx})^{2}>4(eV)^{2}$ then Eq.(\ref{Gamma}) gives $\Gamma ^{tr}\overset{%
\geq }{\sim }(200-300)$ $K$. There is no apparent reason that the intra-band
scattering is very different from the inter-band one (contrary to %
\textcolor{black}{MgB$_{2}$}), then one obtains $\Gamma _{hh}\sim
\Gamma _{ee}\sim \Gamma _{he}\sim \Gamma _{eh}\overset{>}{\sim
}\Gamma ^{tr}/2\sim (100-150)K>>T_{c}$. Since $\Gamma _{hh}\sim
\Gamma _{ee}\gtrsim T_{c}$ then
the intra-band impurity scattering is strongly pair-breaking \cite%
{Kulic-review}, which means that \textit{any unconventional intra-band
pairing is immediately destroyed}. This is also supported by the recent
ARPES measurements on single crystals of %
\textcolor{black}{NdFeAsO$_{0.9}$F$_{0.1}$ }(with $T_{c}=53$ $K$)
where a slightly anisotropic gap around the $\Gamma $ point with
$\Delta \approx 15$ $meV$ was observed. A similar conclusion holds
also for the case $\Gamma _{h}^{tr}\neq \Gamma _{e}^{tr}$ since
for $\omega _{pl,h}^{2}\sim \omega _{pl,e}^{2}$ the resistivity is
dominated by the smaller value of $\Gamma _{h}^{tr}$, $\Gamma
_{e}^{tr}$, i.e. by $\min (\Gamma _{h}^{tr},\Gamma _{e}^{tr})$.

The $s_{\pm }$ pairing is unaffected by the intra-band impurity ($u_{ii}$)
scattering in the Born limit ($N_{i}u_{ij}\ll 1$) \cite{Golubov-Mazin}, \cite%
{Dolgov-Kulic-anis-imp}, \cite{Senga} but it is \textit{sensitive} to the
\textit{inter-band} one $\Gamma _{he}$, $\Gamma _{eh}$, which are
pair-breaking. The pairing interaction in the weak coupling BCS limit is
given by $H=H_{0}+H_{BCS}+H_{imp}$
\begin{equation*}
H_{BCS}=-\sum_{ij}V_{ij}\int dx\psi _{i\uparrow }^{\dag }\psi _{i\downarrow
}^{\dag }\psi _{j\downarrow }\psi _{j\uparrow },
\end{equation*}%
\begin{equation}
H_{imp}=\sum_{ij\sigma =\uparrow ,\downarrow }u_{ij}\int dx\psi _{i\sigma
}^{\dag }\psi _{j\sigma }+c.c.,  \label{Hamilt}
\end{equation}%
where the pairing interaction $V_{ij}$ ($i,j=1,2$, i.e. $h,e$) can be
positive for attraction, negative for repulsion. The self-consistent
equations for the order parameters $\Delta _{i}$ ($i=1,2$, i.e. $h,e$) are
\begin{equation}
\Delta _{i}=\pi T\sum_{j,n}^{-\omega _{cj}<\omega _{n}<\omega _{cj}}\lambda
_{ij}\frac{\tilde{\Delta}_{jn}}{\sqrt{\tilde{\omega}_{jn}^{2}+\tilde{\Delta}%
_{jn}^{2}}}  \label{Delta-i}
\end{equation}%
where $\omega _{cj}$ is the energy cutoff in the $j$-th band and the intra-
and inter-band coupling constants are $\lambda _{ij}=N_{j}(0)V_{ij}$. To
illustrate the effect of the\ multiple impurity scattering beyond the Born
limit we consider the case \textit{of an arbitrary strength of the
inter-band potential} $u_{12}=u_{21}=u$, while $u_{ii}=0$ \cite%
{Dolgov-Kulic-anis-imp}. The renormalized Matsubara frequency $\tilde{\omega}%
_{jn}$ and the gap $\tilde{\Delta}_{jn}$ are given

\begin{equation}
\tilde{\omega}_{1n}=\omega _{n}+\Gamma _{1}\sigma \frac{(\sigma -1)\tilde{%
\delta}_{1n}^{2}\tilde{\omega}_{2n}-\sigma \tilde{\omega}_{1n}\tilde{\delta}%
_{1n}\tilde{\delta}_{2n}}{D_{1n}}  \label{omega}
\end{equation}

\begin{equation}
\tilde{\Delta}_{1n}=\Delta _{1}+\Gamma _{1}\sigma \frac{(\sigma -1)\tilde{%
\delta}_{1n}^{2}\tilde{\Delta}_{2n}-\sigma \tilde{\Delta}_{1n}\tilde{\delta}%
_{1n}\tilde{\delta}_{2n}}{D_{1n}},  \label{Delta}
\end{equation}%
where $D_{1n}=2(\sigma -1)\sigma \tilde{\delta}_{1n}\left( \tilde{\Delta}%
_{1n}\tilde{\Delta}_{2n}+\tilde{\omega}_{1n}\tilde{\omega}_{2n}\right) -%
\left[ 2(\sigma -1)\sigma +1\right] \tilde{\delta}_{1n}^{2}\tilde{\delta}%
_{2n}$ and $\tilde{\delta}_{jn}^{2}=\tilde{\omega}_{jn}^{2}+\tilde{\Delta}%
_{jn}^{2}$. The equations for $\tilde{\omega}_{2n}$, $\tilde{\Delta}_{2n}$
are obtained by replacing $1\longleftrightarrow 2$. The unitary amplitude is
$\Gamma _{i}=c_{imp}/\pi N_{i}(0)$ and $\sigma =\pi
^{2}N_{1}(0)N_{2}(0)u^{2}/(1+\pi ^{2}N_{1}(0)N_{2}(0)u^{2})$. In the unitary
limit one has $N_{i}(0)u\gg 1$ and $\sigma \rightarrow 1$. To illustrate the
pair-breaking effect of the inter-band scattering let us consider the Born
limit of Eqs.(\ref{Delta-i}-\ref{Delta}) , i.e. $N_{i}u<<1$, $\sigma \ll 1$.
For small $c_{imp}$ one has $\Gamma _{he}(\equiv \Gamma _{1})$,$\Gamma
_{eh}(\equiv \Gamma _{2})<<T_{c0}$, where $T_{c0}$ stands for the clean
system. The suppression of $T_{c}$ is given by $(\delta T_{c}/T_{c0})\approx
(\pi \Gamma _{he}\left( \Delta _{h}-\Delta _{e}\right) \left( N_{e}\Delta
_{h}-N_{h}\Delta _{e}\right) /8T_{c0}N_{e}\left( \Delta _{h}^{2}+\Delta
_{e}^{2}\right) )$, where $(\Gamma _{he}/\Gamma _{eh})=N_{e}/N_{h}$. In the
weak coupling limit one has $\left\vert \Delta _{h}/\Delta _{e}\right\vert
\approx \sqrt{N_{e}/N_{h}}$, which gives $\left\vert \Delta _{h}/\Delta
_{e}\right\vert \approx 1$ and $\delta T_{c}\approx -\Gamma _{he}$. We see
that $s_{\pm }$ pairing is affected by the inter-band impurity scattering
only and the pair-breaking of the inter-band scattering is maximal for $%
\Delta _{h}\approx -\Delta _{e}$. This result holds also in the presence of
the intra-band scattering $u_{ii}\neq 0$ in the Born limit, since $u_{ii}$
drops out from equations -- the Anderson theorem. For $\Gamma _{he}\sim
T_{c0}$ then $T_{c}$ is drastically reduced ($T_{c}<<T_{c0}$) in the Born
limit $\sigma \ll 1$. The above estimate $\Gamma _{he}\approx 100-150$ $K$
means that $\Gamma _{he}>T_{c0}$ for the reported single crystals with $%
T_{c}>30$ $K$. At first glance a large $\Gamma _{he}(>T_{c0})$ destroys the $%
s_{\pm }$ pairing. However, the theory, contained in Eqs.(\ref{Delta-i}-\ref%
{Delta}), predicts that in the unitary limit $\pi N_{e}u_{he}\gg
1$ the effect of impurities on $s_{\pm }$ pairing disappears,
i.e.\ the Anderson theorem is restored
\cite{Dolgov-Kulic-anis-imp}. This also holds in the presence of
the intra-band scattering $u_{ii}=\alpha u_{he}$ with $\alpha \neq
1$ \cite{Dolgov-Kulic-anis-imp}. In the unitary limit with $\alpha
=1$ the inter-band pair-breaking effect is strongly amplified by
the intra-band one and $T_{c}$ is strongly suppressed
\cite{Dolgov-Kulic-anis-imp}. If one assumes the Born limit one
has $\Gamma _{he}\sim \pi c_{imp}N_{e}u_{he}^{2}\sim (100-150)K$,
and for the DFT-LDA value $N_{e}\sim 0.6$
\textcolor{black}{states/eV$\cdot$spin} \cite{Mazin-interband} the
Born
limit would be realized in the reported single crystals of %
\textcolor{black}{Ba$_{1-x}$K$_{x}$Fe$_{2}$As$_{2}$} \cite%
{Luo-resistivity-AKFe2As2} for $c_{imp}>10\%$ if $N_{e}u_{he}<0.3$. However,
for a realistic value $c_{imp}\gtrsim 1\%$ one has $\pi N_{e}u_{he}\lesssim 1
$, $\sigma <0.5$ and the impurity scattering in the single crystals of %
\textcolor{black}{Ba$_{1-x}$K$_{x}$Fe$_{2}$As$_{2}$} \cite%
{Luo-resistivity-AKFe2As2} is in the intermediate regime. In this case the
inter-band impurity scattering suppresses $s_{\pm }$ pairing moderately.
However, due to many-body effect the unscreened plasma-frequencies are
renormalized, i.e $\tilde{\omega}_{pl}<\Omega _{pl}$. Optical data analyzed
in the single-band model yield $1.5$ $eV\leq \tilde{\omega}_{pl}<2$ $eV$
\cite{Drechsler-optics}, thus reducing all $\Gamma _{ij}$ by a factor of
about $1.7$, i.e.\ the renormalized value $\Gamma _{ij}^{r}\approx (60-90)$ $%
K$. Strong coupling effects further reduce $\Gamma _{ij}^{r}$ (e.g., in the
single-band case by the factor $1+\lambda $). Nevertheless, the intra-band
scattering $\Gamma _{ii}^{r}$ is sufficiently strong to destroy the
intra-band unconventional pairing. At the same time the inter-band impurity
scattering $\Gamma _{ij}^{r}(>T_{c0})$ is pushed toward the Born limit,
i.e.\ $\sigma <0.5$ thus being more \textit{detrimental} for $s_{\pm }$
pairing than in the unitary limit \cite{Golubov-Mazin}, \cite%
{Dolgov-Kulic-anis-imp}.

The ARPES study of the over-doped %
\textcolor{black}{Ba$_{1-x}$K$_{x}$Fe$_{2}$As$_{2}$}, with $x=1$ and small $%
T_{c}=3$ $K$, shows the absence of the hole band, thus destroying the inter-
(in the hole band) and intra-band pairings \cite{Sato-ARPES-overdoped}. Is
the inter-band interaction $\lambda _{he}$ (and $\lambda _{eh}$) repulsive
or attractive in the Fe-pnictides? This cannot be extracted from resistivity
measurements. However, the \textit{large Fe-isotope effect} on the
polycrystalline samples of \textcolor{black}{SmFeAsO$_{1-x}$F$_{x}$} and %
\textcolor{black}{Ba$_{1-x}$K$_{x}$Fe$_{2}$As$_{2}$} with $\alpha
_{Fe}\approx 0.4$ \cite{Liu-isotope}, if confirmed, implies that
the inter-band pairing is dominated by the attractive EPI. In that
case, the \textit{s-wave }($s_{+} $) pairing is realized in the
hole- and the electron-band with $sign(\Delta _{h}/\Delta
_{e})=+$. The magnetic interaction in this scenario is
pair-weakening, i.e. it is detrimental. If $\alpha _{Fe}$ is going
to be small, then the EEI electron-boson pairing might be
operative. That the repulsive part of the inter-band interaction
due to spin fluctuations might
be small comes out from the recent NMR studies of the $^{75}$%
\textcolor{black}{As} relaxation rate $(T_{1}T)^{-1}$ on %
\textcolor{black}{LaFeAsO$_{1-x}$F$_{x}$} ($0.04\leq x\leq 0.14$)
as a function of $x$ \cite{Nakai-NMR-ASF}. While$(T_{1}T)^{-1}$,
which is dominated by the antiferromagnetic fluctuations,
decreases by increasing doping by almost two orders of magnitude,
$T_{c}$ is practically unchanged. The latter suggests weakness of
the inter-band interaction due to spin
fluctuations. It is worth to mention that the residual in-plane resistivity $%
\rho _{imp}^{ab}$ of \textcolor{black}{(Ba,K)Fe$_{2}$As$_{2}$} in
two different single crystals was different, $70$ and $100$ $\mu
\Omega cm$ respectively, giving $\Gamma ^{tr}\approx (300-600)$
$K$, but $T_{c}$ was unchanged, i.e.\ $T_{c}\approx 28$ $K$
\cite{Yuan-Hc2}. This result is an additional evidence for the
robustness against impurities and the conventional intra-band SC
in these materials. The conventional pairing scenario is also
supported by the data for $H_{c2}(T)$ near $T_{c}$ as a
function of enhanced disorder provided e.g.\ by As vacancies \cite{Fuchs-Hc2}%
, where as a result of disorder the slope of $H_{c2}(T)$ increases, while $%
T_{c}$ is even increased in some samples or remains nearly unchanged.
Similar effects have been observed also in Ref.\cite{Singh-Hc2}. To the best
of our knowledge, only within conventional $s$-wave intra-band pairing the
upper critical field $H_{c2}(T)$ near $T_{c}$ can be readily strongly
enhanced due to the reduction of the well-known spin independent orbital
pair-breaking effect caused by the Lorentz force on Cooper pairs. Within a
dominant $s_{\pm }$ inter-band scenario there is no room for such reduction
effects. The most what can be "explained" by unconventional scenarios is the
robustness of $T_{c}$ in adopting simply an ineffective impurity scattering
mechanism (ascribed to the different orbitals involved in each of the two
bands) \cite{Senga} ; but by no means the reported improved $H_{c2}$ and $%
T_{c}$.

\section{\textbf{Nonmagnetic ($s_{+}$) or magnetic ($s_{\pm }$) pairing}}

The STM conductance \cite{Tesanovic-STM} gives evidence for a single gap,
while in ARPES spectra three gaps were observed - two hole gaps $\Delta
_{h,1}\approx 12$ $meV$ and $\Delta _{h,2}\approx 6$ $meV$ and one electron
gap $\Delta _{e}\approx 12$ $meV$ \cite{epl-ARPES}. Neglecting $\Delta
_{h,2} $ a two-band model (with $\Delta _{h,1}\equiv \Delta _{1}$ and $%
\Delta _{e}\equiv \Delta _{2}$) is applicable. Let us analyze the
Fe-pnictides in the weak-coupling limit and assume that the couplings $%
\lambda _{11}\approx $ $\lambda _{22}\sim \lambda _{12}\sim \lambda _{21}$.
In that case one has for $0<T\leq T_{c}$
\begin{equation}
\frac{\lambda _{\max }-\lambda _{22}}{\lambda _{21}}\approx \frac{\Delta _{1}%
}{\Delta _{2}}\approx \frac{\lambda _{12}}{\lambda _{\max }-\lambda _{11}}%
\approx \sqrt{\frac{N_{2}(0)}{N_{1}(0)}}sign(\lambda _{12}),  \label{ratio}
\end{equation}%
where $\lambda _{\max }=(\lambda _{11}+\lambda
_{22}+\sqrt{(\lambda _{11}-\lambda _{22})^{2}+4\lambda
_{12}\lambda _{21}})/2$ is the maximal eigen-value of the matrix
$\lambda _{ij}$ which determines $T_{c}=1.13\omega _{c}\exp
\{-1/\lambda _{\max }\}$. It gives $(\Delta _{h}/\Delta
_{e})\approx sign(\lambda _{he})$. Note, the couplings $\lambda
^{\prime }$s \ are \textit{effective ones} due to the presence of
attractive and repulsive interactions. So, for $\lambda _{he}>0$
the $s_{+}$ pairing is realized, while for $\lambda _{he}<0$
\textcolor{black}{the} $s_{\pm }$ pairing \textcolor{black}{is
more stable}. The first case is dominated by the conventional EBI
pairing, while the latter one by a repulsive interaction. The
obtained ratio $\left\vert \Delta _{h}/\Delta _{e}\right\vert $ is
similar to the experimental value $\left\vert \Delta _{h}/\Delta
_{e}\right\vert \approx 1$ \cite{epl-ARPES}. This result suggests
that the intra- and inter-band couplings are of the same order. In
this case the specific heat jump at $T_{c}$ is given by $(\Delta
C_{S}/C_{N})\approx
1.43[(N_{h}+N_{e}\delta ^{2})^{2}/(N_{h}+N_{e})(N_{h}+N_{e}\delta ^{4})]%
\overset{<}{\sim }1.43$, with $\delta =\Delta _{e}/\Delta _{h}$. The above
analysis is based on the weak coupling limit and it serves for obtaining a
qualitative picture only, while a full quantitative analysis should be based
on the Eliashberg (strong coupling) theory which is important even in the
weak coupling limit.

\section{\textbf{Discussion}}

The available experimental data do not allow us to extract the intra- and
inter-band couplings and their signs. But some qualitative estimates of $%
\lambda _{11}$, $\lambda _{22}$, $\lambda _{12}$ and $\lambda _{21}$ (note $%
1\equiv h$, $2\equiv e$) are possible in the weak coupling limit. In the
over-doped $Ba_{1-x}K_{x}Fe_{2}As_{2}$, with $x=1$ and $T_{c}^{over}=3$ $K$
\cite{Sato-ARPES-overdoped}, the hole band is absent and $\lambda
_{22}=\lambda _{12}=\lambda _{21}=0$, while in the optimally doped system $%
T_{c}^{opt}\approx 40$ $K$ one expects $\lambda _{11}\approx \lambda
_{22}\sim \lambda _{12}\approx \lambda _{21}$. From these data one obtains $%
\omega _{c}\approx 260$ $K$, $\lambda _{11}\approx \lambda _{22}\approx 0.22$
and $\lambda _{12}\approx \lambda _{21}\approx 0.28$. The Debye temperature
in the Fe-pnictides is $\omega _{D}\approx 280$ $K$($\approx \omega _{c}$)
and the magnitudes of the $\lambda ^{\prime }$s are in the range of the LDA
value for the EPI coupling constant. By approaching the overdoped region one
expects that $\lambda _{12}$ and $\lambda _{22}$ decreases and vanishes, as
it is observed in ARPES spectra of %
\textcolor{black}{Ba$_{1-x}$K$_{x}$Fe$_{2}$As$_{2}$}. This
scenario could be tested in clean systems by measuring the
collective Leggett mode which is
specific for clean two\textcolor{black}{(or multi)}-band) systems with %
\textcolor{black}{an} internal inter-band Josephson effect. In
this case the contribution to the total energy, for small $\lambda
_{12},\lambda _{21}$, is $E_{inter}=-E_{0}\cos (\theta _{1}-\theta
_{2})$. For $s_{+}$ pairing with $\lambda _{12}(\lambda _{21})>0$
one has $E_{0}>0$ and the ground state
is realized for $\theta _{1}-\theta _{2}=0$ - the $0$-contact, while for $%
s_{\pm }$ pairing $E_{0}<0$ and $\theta _{1}-\theta _{2}=\pi $ - the $\pi $%
-contact. In both cases the relative phase ($\theta _{1}-\theta _{2}$) of
the two order parameters ($\Delta _{1,2}=\left\vert \Delta _{1,2}\right\vert
\exp \{-i\theta _{1,2}\}$) oscillates with the Leggett-frequency
\begin{equation}
\omega _{L}^{2}=\frac{4\left\vert \lambda _{12}+\lambda _{21}\right\vert
\left\vert \Delta _{1}\Delta _{2}\right\vert }{\lambda _{11}\lambda
_{22}-\lambda _{12}\lambda _{21}}.  \label{L-mode}
\end{equation}%
The condition for the existence of the Leggett mode is $D\equiv \lambda
_{11}\lambda _{22}-\lambda _{12}\lambda _{21}>0$ and it is undamped for $%
\omega _{L}<2\left\vert \Delta _{1}\right\vert $. This can be
realized in low doped $Fe$-pnictides where it is expected that
$\left\vert \lambda _{12}\right\vert \ll \lambda _{11}$ and $D>0$,
while by doping toward the optimal doping it may disappear or
become overdamped. This proposal can be tested
\textcolor{black}{by studying the corresponding} tunnelling and
electronic Raman spectra - like in \textcolor{black}{MgB$_{2}$} \cite%
{Ponomarev-Maksimov}.

What is the possible origin of the EBI pairing in Fe-pnictides?
The isotope effect, if confirmed, points to the importance of EPI.
However, the effective Coulomb interaction $\mu ^{\ast }$ (see
below) suppresses EPI and in this case an excitonic pairing (EEI)
might be helpful \textcolor{black}{to compensate} the Coulomb
interaction. In Fe-pnictides the As ions posses large
polarizability $\alpha _{e}=(9-12)$ \AA $^{3}$ \cite{Sawatzky}
which can partly compensate or overcompensate the Coulomb
interaction, thus
changing the electronic dielectric function $\varepsilon _{el}(\mathbf{q}%
,\omega =0)=\varepsilon _{c}(\mathbf{q},0)+\varepsilon _{ex}(\mathbf{q},0)$,
where $\varepsilon _{c}$ is the dielectric function of conduction carriers.
Since in a multi-band model this mechanism can be obscured by numerous
parameters we elucidate it in the single-band model by assuming that the
effective pairing interaction $V_{pair}(\xi ,\xi ^{\prime })$ contains three
dominant parts, two attractive - EPI with the potential $V_{epi}$ and EEI
with $V_{eei}$, and one repulsive with $V_{c}$
\begin{equation}
V_{pair}(\xi ,\xi ^{\prime })=V_{epi}(\xi ,\xi ^{\prime })+V_{eei}(\xi ,\xi
^{\prime })-V_{c}(\xi ,\xi ^{\prime }),  \label{Vpair}
\end{equation}%
where $V_{epi}(\xi ,\xi ^{\prime })=V_{epi}>0$ for $\left\vert \xi
\right\vert ,\left\vert \xi ^{\prime }\right\vert \leqslant \omega _{ph}$; $%
V_{eei}(\xi ,\xi ^{\prime })=V_{eei}>0$ for $\left\vert \xi \right\vert
,\left\vert \xi ^{\prime }\right\vert \leqslant \omega _{ex}$ and $V_{c}(\xi
,\xi ^{\prime })=V_{c}>0$ $\left\vert \xi \right\vert ,\left\vert \xi
^{\prime }\right\vert \leqslant \omega _{c}$. The characteristic phonon,
exciton and Coulomb energies are related by $\omega _{ph}\ll \omega
_{ex},\omega _{c}$. In the limit when the effective exciton and Coulomb
interactions are small, i.e. when $\lambda _{ex}^{\ast }\mu ^{\ast }\ll
\lambda _{ex}^{\ast },\mu ^{\ast }$, the critical temperature is given by
\begin{equation}
T_{c}\approx 1.14\omega _{ph}\exp \{-1/(\lambda _{ep}-\mu ^{\ast }+\lambda
_{ex}^{\ast }\},  \label{Tc}
\end{equation}%
where $\mu ^{\ast }=\mu /(1+\mu \ln (\omega _{c}/\omega _{ph}))$, and $%
\lambda _{ex}^{\ast }=\lambda _{ex}/(1-\lambda _{ex}\ln (\omega _{ex}/\omega
_{ph}))$. The corresponding coupling constants are $\lambda
_{ep}=N(0)V_{epi} $, $\lambda _{ex}=N(0)V_{eei}$ and $\mu =N(0)V_{c}$. For $%
\lambda _{ex}<\mu $ the exciton pairing may compensate the Coulomb repulsion
giving $\lambda _{ex}^{\ast }>\mu ^{\ast }$. In the single band case it is
difficult to realize an excitonic mechanism of pairing since the electronic
(not the total!) dielectric function $\varepsilon _{el}(q,\omega =0)$ must
be positive ($\varepsilon _{el}(q,0)>0$) in order to keep the lattice
stability \cite{Kulic-review}. In a multi-band case the physics can be
different and the restrictions for an excitonic mechanism can be relaxed.

\section{\textbf{Conclusions}}

The recent experimental results in single crystals of the Fe-pnictides give
evidence for: (\textit{i}) the multi-band character of pairing; (\textit{ii}%
) the presence of a large amount of nonmagnetic impurities (and other
defects) which are strong pair-breakers for unconventional pairing. The
result (ii) strongly favors \textit{conventional nodeless intra-band pairings%
} with $\lambda _{hh}$,$\lambda _{ee}>0$, which are due to a non-magnetic
electron-boson interaction (EBI), either to phonons or eventually to
excitons, or both are operative. If the large Fe-isotope effect $\alpha
_{Fe}\approx 0.4$ is confirmed, then EPI is favored. The EBI inter-band
coupling with $\lambda _{he}(\lambda _{eh})>0$ favors $s_{+}$ pairing with $%
sign(\Delta _{h}/\Delta _{e})=+$. The $s_{\pm }$ pairing with $sign(\Delta
_{h}/\Delta _{e})=-$ is favored by repulsive inter-band interactions with $%
\lambda _{he}<0$. This pairing is less favorable due to the detrimental
effect of disorder present in single crystals and polycrysta%
\textcolor{black}{l}s and due to EBI scattering. Even, if
excitonic effects would not give solely a high $T_{c}$ they may
nevertheless strengthen a pairing due to EPI.

\centerline{***} We thank Zlatko Te\v{s}anovi\'{c} for useful discussions.



\begin{thebibliography}{99}
\bibitem{Senga}
\textcolor{black}{For the vicinity of the very special 'unitary
limit' point in the scattering parameter space the authors
\cite{Dolgov-Kulic-anis-imp} find a minor suppresion of the
$s_{\pm}$ pairing by the interband impurity scattering. Note that
even in this special case with preserved $s_{\pm}$ symmetry of the
order parameter in the presence of dominant repulsive interband
interaction, our main statement of $s$-wave intraband scattering
being relevant for the enhanced slope of $H_{c2}(T)$ and the
resistivity in the normal state does hold. In this case $T_c$
might be considerably enhanced by the repulsive interband
scattering.}
\end{thebibliography}
\end{document}